\documentclass{article}[11pt]
\usepackage{custom}

\usepackage{todonotes}   
\setuptodonotes{fancyline, color=blue!10, size=\small}

\title{Letter of Intent: \\
AICE - Atom Interferometer CERN Experiment}
 
\author[1]{Charles~Baynham,}
\affiliation[1]{High Energy Physics Group, Blackett Laboratory, Imperial College, Prince Consort Road, London, SW7 2AZ, UK}

\author[2]{Andrea~Bertoldi,}
\affiliation[2]{IOGS, LP2N, Université Bordeaux, CNRS, UMR 5298, F-33400 Talence, France}

\author[3]{Diego~Blas,}
\affiliation[3]{Institut de F\'{i}sica d’Altes Energies (IFAE), The Barcelona Institute of Science and Technology, Campus UAB, 08193 Bellaterra (Barcelona), Spain;
Instituci\'{o} Catalana de Recerca i Estudis Avan\c{c}ats (ICREA),
Passeig Llu\'{i}s Companys 23, 08010 Barcelona, Spain}

\author[1,4]{Oliver~Buchmueller$^*$,}
\affiliation[4]{University of Oxford, South Parks Road, Oxford OX1 3PU, UK}

\author[5]{Sergio~Calatroni,}
\affiliation[5]{CERN, 1211 Geneva 23, Switzerland}

\author[6]{Vassilis~Charmandaris,}
\affiliation[6]{Chairman, Board of Directors, Foundation for Research and Technology – Hellas (FORTH), Vassilika Vouton, 70013 Heraklion, Creta, Greece}

\author[7]{Maria Luisa (Marilù)~Chiofalo,}
\affiliation[7]{Department of Physics, University of Pisa, Largo Bruno Pontecorvo 3 56126 Pisa, Italy; INFN-Pisa, Largo Bruno Pontecorvo 3 56126 Pisa, Italy}

\author[8]{Pierre Cladé,}
\affiliation[8]{Laboratoire Kastler Brossel, Sorbonne Universit{\'{e}} PSL, Coll{\`{e}}ge de France, 75005 Paris, France}

\author[9]{Jonathon~Coleman,}
\affiliation[9]{Department of Physics, University of Liverpool, Merseyside, L69 7ZE, UK}

\author[10]{Fabio~Di~Pumpo,}
\affiliation[10]{Institut f{\"u}r Quantenphysik and Center for Integrated Quantum Science and Technology (IQST), Universit{\"a}t Ulm, Albert-Einstein-Allee 11, 89081 Ulm, Germany}

\author[11]{John~Ellis$^*$,}
\affiliation[11]{Physics Department, King’s College London, Strand, London, WC2R 2LS, UK}

\author[12]{Naceur~Gaaloul,}
\affiliation[12]{Leibniz Universit\"at Hannover, Institut f\"ur Quantenoptik, Welfengarten 1, 30167 Hannover, Germany}

\author[8]{Saïda Guellati-Khelifa,}

\author[13]{Tiffany~Harte,}
\affiliation[13]{Cavendish Laboratory, University of Cambridge, J J Thomson Avenue,
Cambridge, CB3 0US, UK}

\author[1]{Richard~Hobson,}

\author[14]{Michael~Holynski,}
\affiliation[14]{School of Physics and Astronomy, University of Birmingham, B152TT Edgbaston, UK}
 
\author[14,15]{Samuel~Lellouch,}
\affiliation[14]{School of Engineering, University of Birmingham, B152TT Edgbaston,  UK}
 
 \author[16,17]{Lucas~Lombriser,}
\affiliation[16]{Department of Applied Future Technologies, University of Applied Sciences of the Grisons, Pulverm\"uhlestrasse 57, 7000 Chur, Switzerland}
\affiliation[17]{D\'epartement de Physique Th\'eorique, Universit\'e de Gen\`eve, 24 quai Ernest Ansermet, 1211 Gen\`eve 4, Switzerland}

\author[18]{Elias~Lopez~Asamar,}
\affiliation[18]{Departamento de F{\'i}sica Te{\'o}rica, Universidad Autonoma de Madrid, 28049 Madrid, Spain;
Instituto de F{\'i}sica Te{\'o}rica UAM-CSIC, 28049 Madrid, Spain}

\author[17,19]{Michele~Maggiore,}
\affiliation[19]{Gravitational Wave Science Centre, Universit\'e de Gen\`eve, Gen\`eve, Switzerland}

\author[11]{Christopher~McCabe,}

\author[13]{Jeremiah~Mitchell,}

\author[12]{Ernst~M.~Rasel,}

\author[17,19]{Federico~Sanchez~Nieto,}

\author[20]{Wolfgang~Schleich,}
\affiliation[20]{Institut f{\"u}r Quantenphysik and Center for Integrated Quantum Science and Technology (IQST), Universit{\"a}t Ulm, Albert-Einstein-Allee 11, 89081 Ulm, Germany;
Institute for Quantum Science and Engineering (IQSE), and Texas A\&M AgriLife Research and Hagler Institute for Advanced Study, Texas A\&M University, College Station, TX 77843-4242, USA}

\author[12]{Dennis~Schlippert,}

\author[13]{Ulrich~Schneider,}

\author[17,19]{Steven~Schramm,}

\author[21]{Marcelle~Soares-Santos,}
\affiliation[21]{University of Zurich, Winterthurerstrasse 190, 8057 Zurich, Switzerland}

\author[22]{Guglielmo~M.~Tino,}
\affiliation[22]{Dipartimento di Fisica e Astronomia and LENS, Universit\`{a} di Firenze, INFN Sezione di
Firenze, CNR-INO, via Sansone 1, I-50019 Sesto Fiorentino, Italy}

\author[9]{Jonathan~N.~Tinsley,}

\author[23]{Tristan~Valenzuela,}
\affiliation[23]{Rutherford Appleton Laboratory, UKRI-STFC, Harwell Campus, Didcot, OX11 OQX, UK}

\author[24]{Maurits~van~der~Grinten,}
\affiliation[24]{PPD, Harwell Campus, Science and Technologies Facilities Council, UKRI, OX11 0QX, UK }

\author[25]{Wolf~von~Klitzing,}
\affiliation[25]{Foundation for Research and Technology – Hellas (FORTH), Institute of Electronic Structure and Lasers (IESL), Vassilika Vouton, 70013 Heraklion, Creta, Greece}

\date{\today}

\begin{document}

\abstract{We propose a ${\cal O}(100)$m Atom Interferometer (AI) experiment - AICE - to be installed against a wall of the PX46 access shaft to the LHC. AICE would probe unexplored ranges of the possible couplings of bosonic ultralight dark matter (ULDM) to atomic constituents and undertake a pioneering search for gravitational waves (GWs) at frequencies intermediate between those to which existing and planned experiments are sensitive, among other fundamental physics studies. A conceptual feasibility study~\cite{Arduini:2851946} showed that this AI experiment could be isolated from the LHC by installing a shielding wall in the TX46 gallery, and surveyed issues related to the proximity of the LHC machine, 
finding no technical obstacles. A detailed technical implementation study~\cite{Arduini:2025jhe} has shown that the preparatory civil‐engineering work, installation of bespoke radiation shielding, deployment of access‐control systems and safety alarms, and installation of an elevator platform could be carried out during LS3, allowing installation and operation of the AICE detector to proceed during Run~4 without impacting HL-LHC operation. These studies have established that PX46 is a uniquely promising location for an AI experiment. We foresee that, if the CERN management encourages this LoI, a significant fraction of the Terrestrial Very Long Baseline Atom Interferometer (TVLBAI) Proto-Collaboration~\cite{abend_terrestrial_2024,abdalla_terrestrial_2025,TVLBAI3,TVLBAIESPP} may wish to contribute to AICE, as discussed in Appendix A.\\
~~\\
CERN-LHCC-2025-008/LHCC-I-042 \\
~~\\
$^*$ Contact person}

\maketitle

\section{Executive Summary}
Quantum sensors are attracting increasing attention for their potential to make precise measurements within the Standard Model (SM) and search for possible new physics beyond the Standard Model (BSM). Among the proposed applications of Quantum Technology to Fundamental Physics, one of the most interesting is Atom Interferometry (AI), which is based on the superposition and interference of atomic wave packets, and combines state-of-the-art atomic clock technology with established techniques for building inertial sensors. AI offers interesting prospects for probes of ULDM and pioneering searches for GWs in a hitherto unexplored frequency range, as well as other fundamental physics studies~\cite{Badurina:2019hst}. First-generation vertical AI experiments of length ${\cal O}(10)$m have been constructed in Hannover~\cite{schlippert2020matter}, at Stanford~\cite{Overstreet:2021hea} and in Wuhan~\cite{zhou2011development}, a Technical Design Report for a 10m AI experiment at Oxford has been completed~\cite{Bongs:2025rqe}, and ${\cal O}(100)$m vertical and horizontal AI experiments are under construction at Fermilab~\cite{MAGIS-100:2021etm} and in France~\cite{Canuel:2017rrp}, respectively.

CERN is a compelling possible site for a vertical AI of length ${\cal O}(100)$m, in view of its physical and technical infrastructure including LHC access shafts, and its experience in hosting international experimental collaborations. CERN's Physics Beyond Colliders (PBC) programme has provided a suitable framework for exploring the feasibility of installing such an AI at CERN~\cite{Arduini:2851946,Arduini:2025jhe}, which would extend the scope of the CERN experimental programme in exciting new directions aligned with the objectives of CERN's Quantum Technology Initiative (QTI), with no impact on the exploitation of the LHC. We anticipate networking the CERN AI experiment with other ${\cal O}(100)$m AI experiments~\cite{MAGIS-100:2021etm,Canuel:2017rrp} as is currently done by large laser interferometry experiments~\cite{LIGOScientific:2014pky,VIRGO:2014yos,Aso:2013eba}.

\begin{figure}
    \centering
    \includegraphics[width=0.45\linewidth]{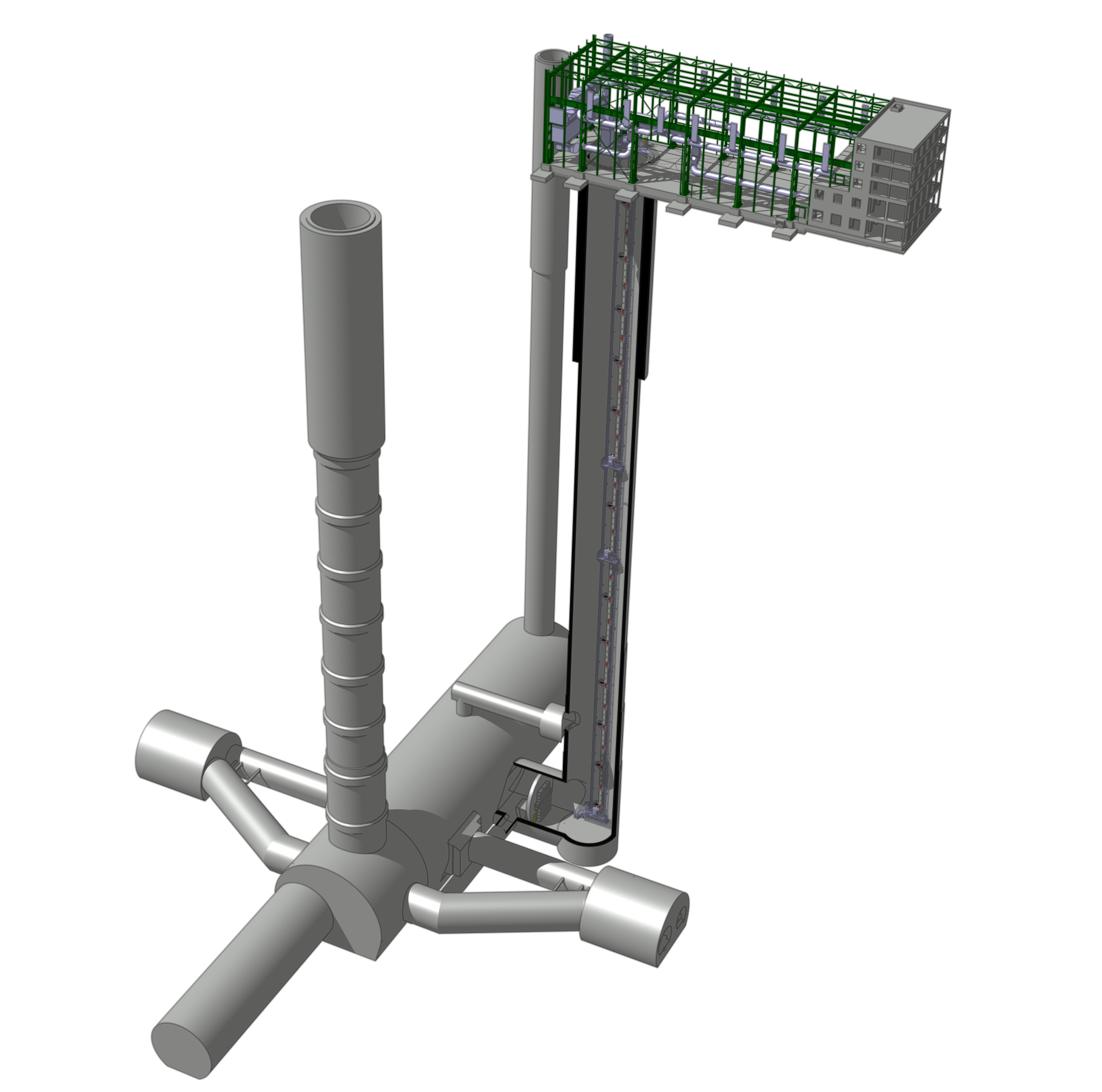}
    \includegraphics[width=0.45\linewidth]{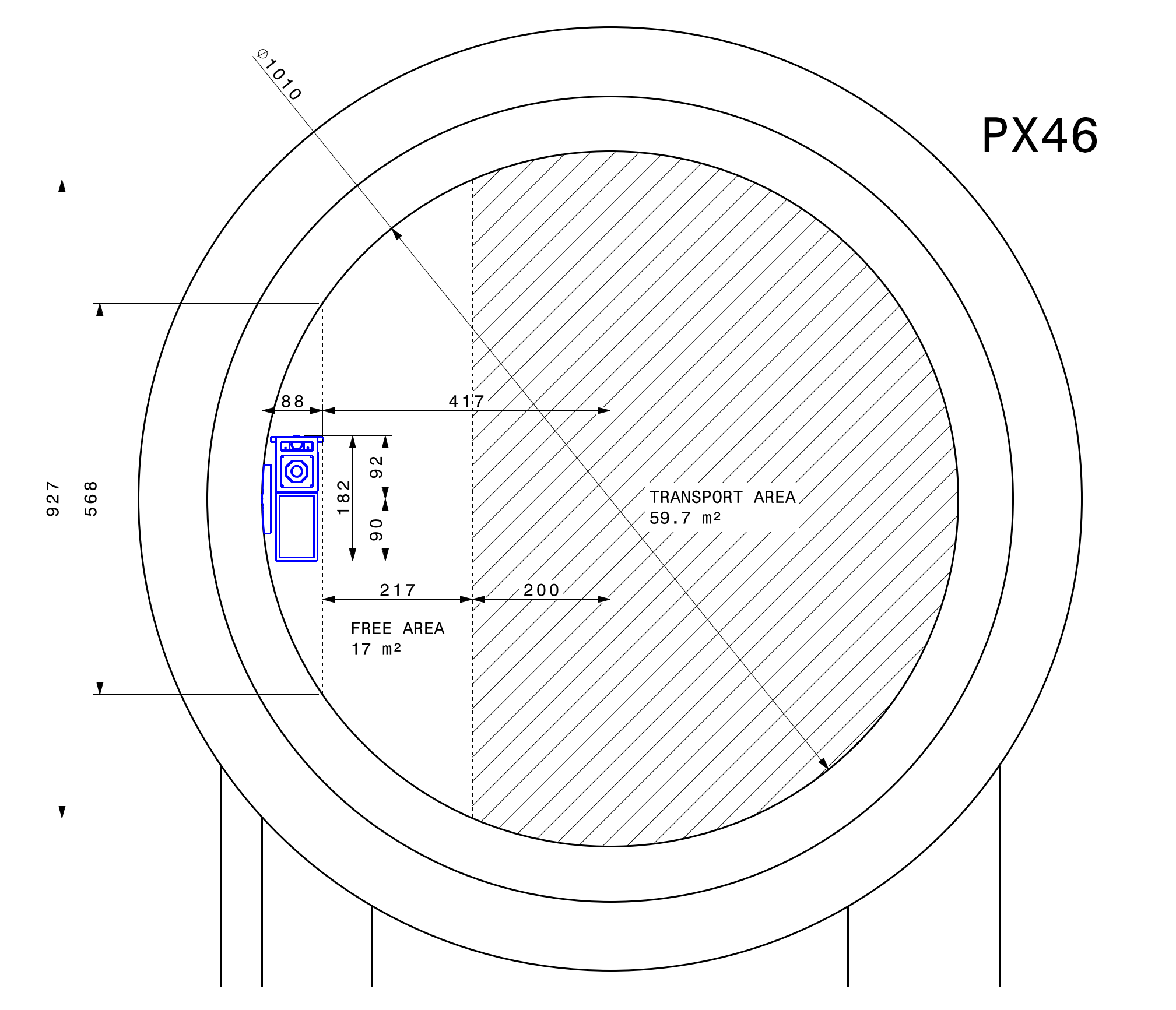}
    \caption{{\it Left panel}: General schematic view of the possible implementation of an AI in the PX46 access shaft~\cite{Arduini:2851946,Arduini:2025jhe}. {\it Right panel}: Horizontal cross section of the PX46 shaft, showing that the proposed location of the AI experiment (blue) is compatible with the space required for transport of LHC components (hatched area).}
    \label{fig:PX46}
\end{figure}

A Conceptual Feasibility Study (CFS)~\cite{Arduini:2851946} has identified the PX46 access shaft to the LHC as the most promising CERN site for a ${\cal O}(100)$m vertical AI that we call AICE (Atom Interferometer CERN Experiment). The left panel of Fig.~\ref{fig:PX46} shows a schematic view of PX46 and the layout of the civil engineering infrastructure at Point 4 on the LHC ring. The vertical height from the SX4 surface building down to the level of the LHC is $\sim 143$m and the internal diameter of the shaft is 10.1m. PX46 provides access to the main LHC radiofrequency (RF) system and its primary use is for occasionally raising and lowering technical equipment. However, as seen in the right  panel of Fig.~\ref{fig:PX46}, a substantial fraction of the horizontal cross-section of the PX46 shaft is not required for LHC access, and would be large enough to accommodate an AI experiment.

As part of the CFS, exploratory seismic measurements were made at the top and bottom of PX46, which were used to estimate the possible level of Gravity Gradient Noise (GGN), an important source of background, which was found to be similar to other sites under consideration~\cite{Arduini:2851946}.  
One issue that is specific to PX46 is that of electromagnetic (EM) noise associated with the LHC RF system and other electrical equipment at depth and on the surface. The EM noise levels have been measured and found not to be of concern. Moreover, the variation in the ambient magnetic field during a ramp of the LHC magnets is small and slow, and can be dealt with by the magnetic shield of the AI~\cite{Arduini:2851946}. Another issue specific to Point 4 of the LHC is the possibility of a major helium release from the LHC cryogenic system. This eventuality was also considered in~\cite{Gaignant:2021,Hakulinen:2022,Arduini:2851946,Arduini:2025jhe}, with the conclusion that existing LHC ventilation and safety systems would be able to deal safely with any such helium release.

An issue for any LHC access shaft is the unlikely possibility of a catastrophic LHC beam loss near the base of the shaft, in view of which radioprotection (RP) measures must be foreseen. These would include in particular the installation of a protective shielding wall in TX46 at the base of PX46, with an access door and provision for temporary opening when LHC equipment must be installed or removed. A detailed, costed solution has been studied~\cite{Arduini:2025jhe} - see Section~3. General safety considerations imply the need for extending the LHC access control system to PX46 and provision for rapid evacuation of experimental personnel when necessary, such as in the event of a fire in the nearby UX45 cavern, which will require an ad hoc elevator system. As also discussed in Section~3, a technical solution has been designed and a cost estimate has been provided~\cite{XLIndustries:2023}.

These conceptual and implementation studies found no showstoppers for siting a $\sim 100$m vertical AI in PX46. The main cost drivers for preparing PX46 for the installation of such an experiment were identified, including the costs of the radioprotection shielding wall and the rapid evacuation system. These preparation costs are small compared to that of the experiment itself - see Section~4. As also discussed in Section~4, a technical timeline has been developed, which indicates that the essential site preparation work could be accomplished during LS3~\cite{LTS:ACC-PM-MS}, enabling the installation and operation of the AI experiment to be undertaken during Run~4 without impacting LHC operations~\cite{Arduini:2025jhe}.



\section{Physics Goals}
The nature of Dark Matter (DM) is one of the greatest puzzles in fundamental physics and astrophysics~\cite{Bertone:2016nfn}, and lies beyond the scope of the Standard Model. The favoured hypotheses include the possibilities that it is composed of non-relativistic particles such as weakly-interacting massive particles (WIMPs) or coherent waves of ultralight bosons. Experiments at the LHC and elsewhere have not yet found any evidence for WIMPs, though searches will continue during Run 3 of the LHC and at the high-luminosity LHC (HL-LHC). However, in the meantime the search for Ultra-Light Dark Matter (ULDM)~\cite{Graham:2015ifn} is attracting growing interest, and this is one of the principal scientific objectives of atom interferometry (AI) experiments such as AION~\cite{Badurina:2019hst}.

The other principal objective of such experiments is the search for GWs in the range of frequencies around 1 Hz that is intermediate between the peak sensitivities of present terrestrial experiments such as LIGO~\cite{LIGOScientific:2014pky}, Virgo~\cite{VIRGO:2014yos} and KAGRA~\cite{Aso:2013eba} (LVK), and the planned space-borne experiment LISA~\cite{LISA:2017pwj}. Among the targets of experiments in this intermediate frequency range are mergers of black holes (BHs) with masses intermediate between those whose mergers have been detected by LIGO and Virgo and the supermassive black holes (SMBHs) detected in the centres of galaxies~\cite{EventHorizonTelescope:2019dse,EventHorizonTelescope:2022wkp}. Detectors in the intermediate frequency range may also be sensitive to a background of GWs produced by fundamental physics processes such as first-order phase transitions in the early Universe or the evolution of a network of cosmic strings~\cite{Badurina:2019hst}.

\begin{figure}[t!]
\centering
\vspace{-0.7cm}
\includegraphics[width=0.4\textwidth]{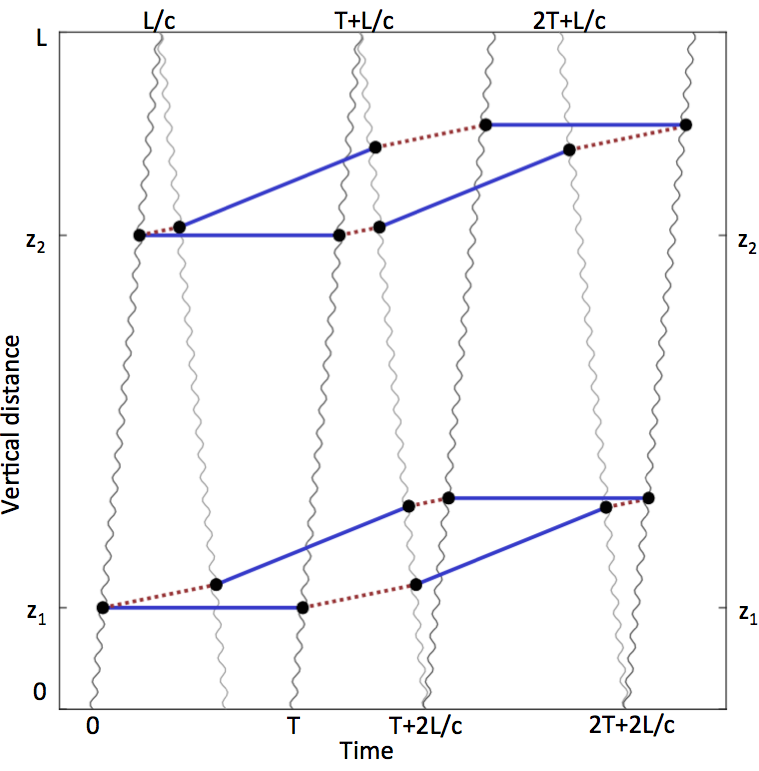}
\includegraphics[width=0.3\textwidth]{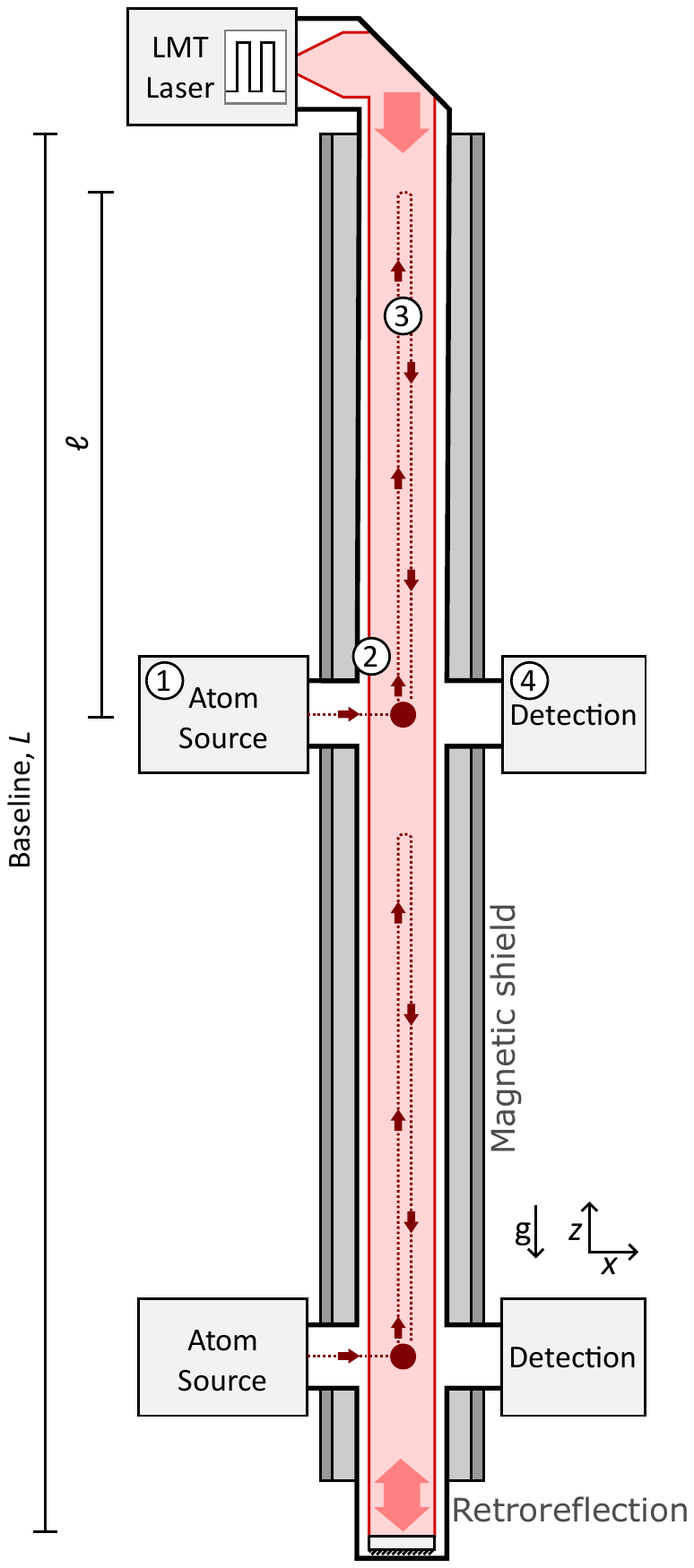}
\caption{{\it Left panel}: Space-time diagram of the operation of a pair of cold-atom interferometers based on single-photon transitions between the ground state (blue) and the excited state (red dashed). Height is shown on the vertical axis and the time axis is horizontal.
The laser pulses (wavy lines) traversing the baseline from opposite ends are used to divide, redirect, and recombine the atomic matter waves, yielding interference patterns that are sensitive to the modulation of the atomic transition frequency caused by coupling to ULDM, or the space-time distortions caused by
GWs. {\it Right panel}: Conceptual scheme of an AI experiment with two atom sources that project clouds vertically, addressed by a single laser source (this diagram is not to scale).}
\label{fig:space-time}
\end{figure}

The basic principle of AI is illustrated in Fig.~\ref{fig:space-time}. It is analogous to that of laser interferometry as employed by current GW detectors. In an atom interferometer, each atom’s wavefunction is coherently split by a laser pulse into a superposition of ground- and excited-state components that follow two diverging trajectories. A subsequent ‘mirror’ pulse exchanges the populations along these paths, and a final pulse recombines them, leading to interference between the two trajectories. 
Each laser interaction with an atom imparts momentum, so the two populations follow different space-time trajectories, as seen in Fig.~\ref{fig:space-time}. Interactions of the atoms with a coherent wave of ULDM may alter the excited atomic energy level, modifying the atomic phase and hence the interference pattern. In practice, AI experiments use two or more sources of atoms that are exposed to the same laser beam, as also shown in Fig.~\ref{fig:space-time}, thereby minimising the effects of laser noise,~\footnote{This has been demonstrated in the laboratory, see, e.g.,~\cite{AION:2025igp}.} and measure the differences between the interference patterns they exhibit, which are sensitive to the space-time dependence of the ULDM field density. Such differential measurements are also sensitive to the distortions of space-time caused by the passage of a GW~\cite{Badurina:2019hst}.

In the AICE experiment proposed here the cold atom clouds are launched vertically into a vacuum tube and follow ballistic trajectories modified by laser pulses that are also directed vertically downwards and reflected in a mirror at the bottom of the tube. This configuration has the advantage that the atom clouds may have a relatively long flight time, $T$, that is limited only by the length, $L$, of the tube. During this time, the atoms may be subjected to multiple laser pulses, $n$, enabling large momentum transfers (LMTs) that enlarge the `hysteresis diamonds’ illustrated in Fig.~\ref{fig:space-time}.~\footnote{The AION Collaboration has demonstrated LMTs using several hundred laser pulses - {\it in preparation}.} Also, separating the sources by relatively large vertical distances, $\Delta z$, enables atom clouds to be projected over large vertical distances and helps mitigate the GGN due to seismic movements in the surrounding rock, which decay exponentially with depth. The indicative values of the experimental parameters assumed here for a two-interferometer design are listed in Table~\ref{tab:AI100parameters}, including also the repetition rate, $\Delta t$, the laser phase noise and the experimental duration, $T_{\rm Int}$, though we anticipate that more ambitious parameters should be attainable, e.g., by increasing $n$, the number of laser pulses and hence the number of momentum transfers, increasing $N$, the number of atoms in each cloud, as well as squeezing the atom states, and that there would be advantages in adding more sources along the vacuum tube~\cite{chaibi2016low,Badurina:2022ngn}. These would help control the impact of GGN, particularly in the presence of different rock strata.

\begin{table}[h]
 \caption{Indicative experimental parameters for a 100~m AI with two cold atom sources.}
 \vspace{3mm}
 \label{tab:AI100parameters}
  \centering
  \begin{tabular}{c|c|c|c|c|c|c|c}
   L [m] & T [s] & $n$ & $\Delta z$ [m] & $N$ & $\Delta t$ [s] & Phase noise [1/$\sqrt{\rm Hz}$] & $T_{\rm{Int}}$ [s] \\
   \hline
   100 & 1.4 & 5494 & 50 & $10^8$ &  1.5 & $10^{-5}$  & $10^8$
  \end{tabular} 
\end{table}

The scheme we propose for a vertical AI has a single laser source located at the top of the vertical vacuum tube, in the existing surface building SX4. One of the atom sources is located at the bottom of the vacuum tube, and one or more additional sources (ultimately 5 to 10) will be installed along the length of the tube, which is surrounded by a magnetic shield. The right panel of Fig.~\ref{fig:space-time} illustrates schematically the layout of an AI with two sources.

\begin{figure*}[h!]
 \centering
 \includegraphics[width=0.48\textwidth]{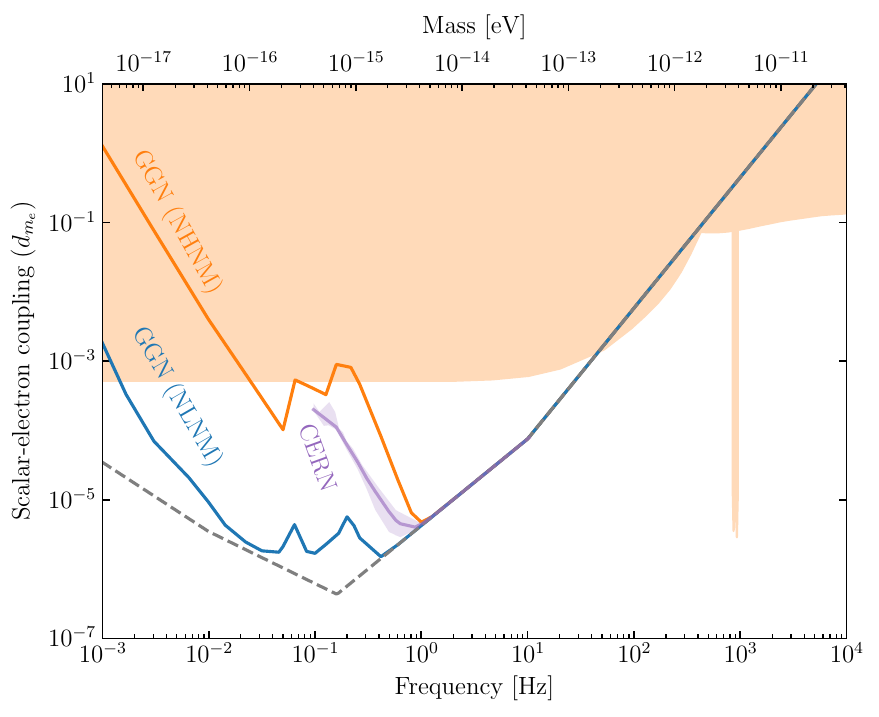}
 \includegraphics[width=0.50\textwidth]{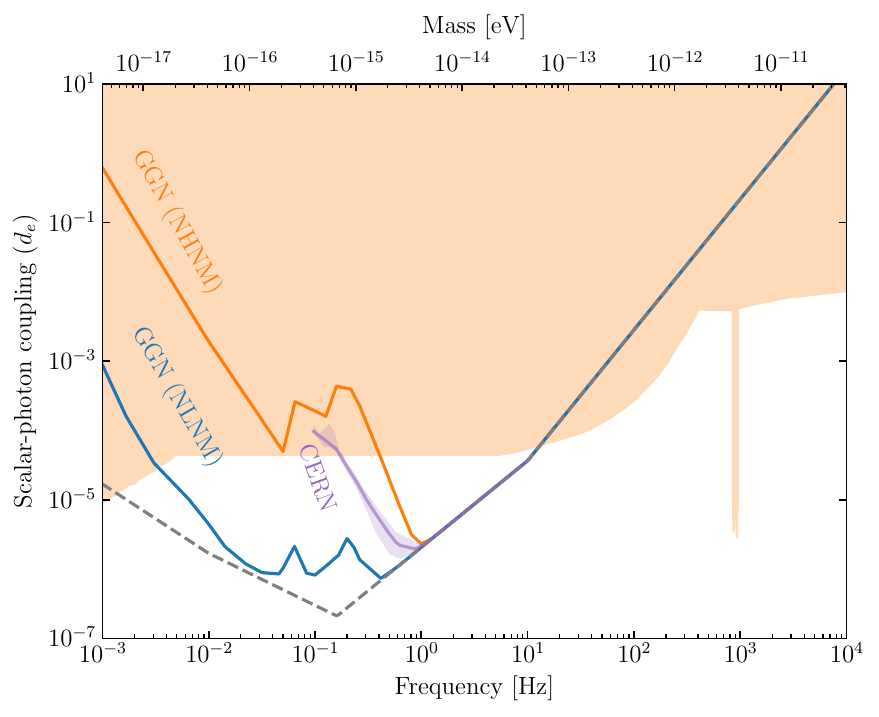}\\
 \includegraphics[width=0.48\textwidth]{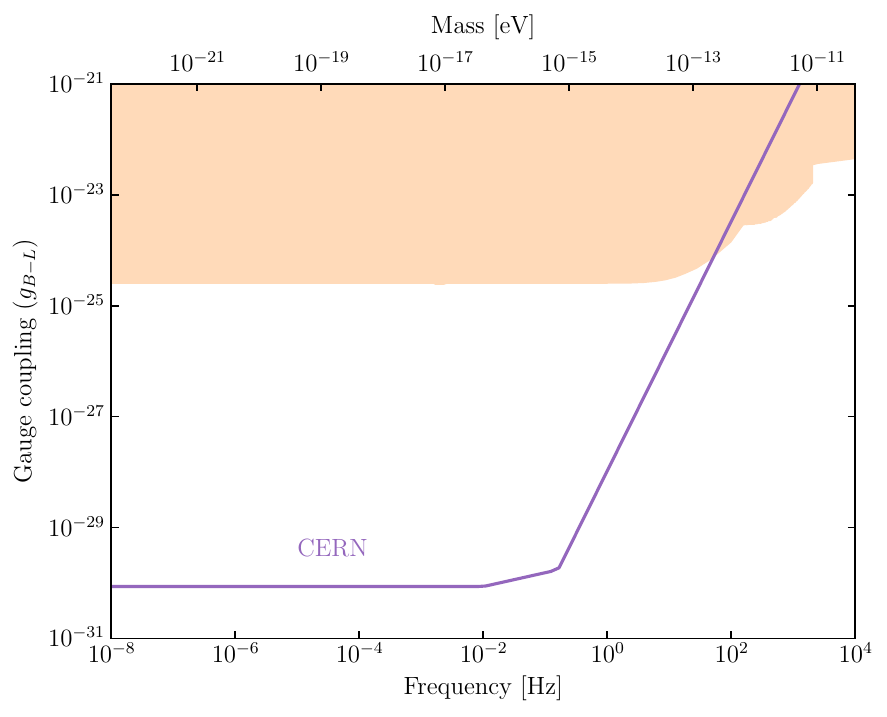}
 \includegraphics[width=0.50\textwidth]{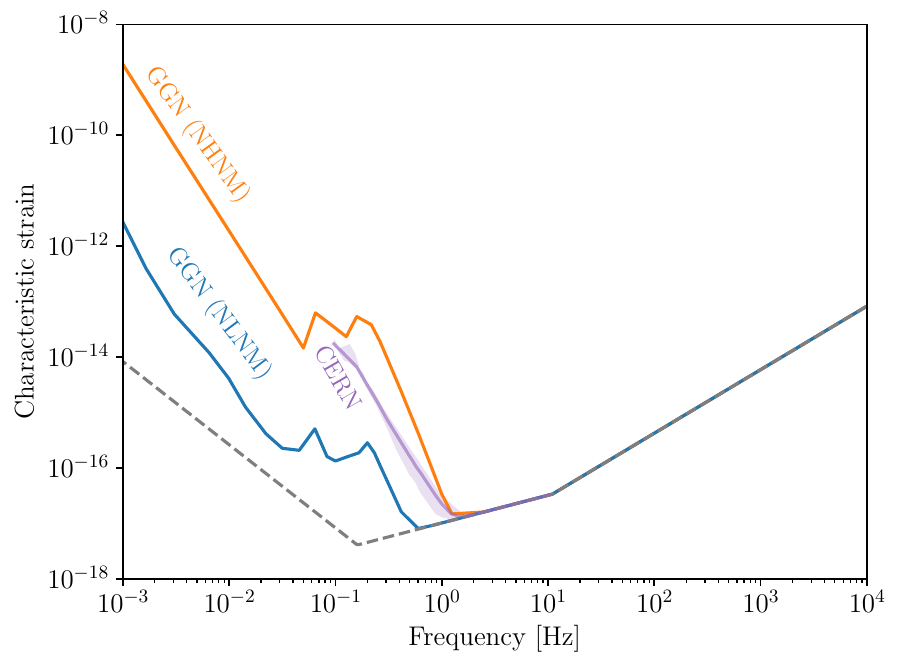}
 \caption{ \label{fig:AI100Sensitivities} Sensitivities of
 a 100-m AI to the couplings of scalar ULDM to the electron ({\it upper left panel}), to the photon ({\it upper right panel}), and to vector dark matter coupled to $B - L$ ({\it lower left panel}). The {\it lower right panel} shows the sensitivity to GW strain. The Atom Shot Noise (ASN) calculated from the indicative experimental parameters in Table~\ref{tab:AI100parameters} is shown by black dashed lines, the new high- (low-)noise model (NHNM) (NLNM) for gravity gradient noise (GGN) is indicated by solid orange (blue) lines. The GGN calculated on the basis of seismic measurements at the PX46 site is indicated by violet lines, with shading corresponding to the diurnal fluctuations in vertical surface motion reported in~\cite{Arduini:2851946}.
 The surrounding rock is assumed to have properties similar to sandstone (molasse). Plots taken from~\cite{Arduini:2851946}.}
\end{figure*}

Figure~\ref{fig:AI100Sensitivities} illustrates the capabilities of a 100-m vertical AI detector assuming the indicative experimental parameters listed in Table~\ref{tab:AI100parameters}, and broadband operation. The upper panels illustrate the sensitivities of such an AI experiment to scalar ULDM~\cite{Arvanitaki:2016fyj,Badurina:2019hst}, see also~\cite{Badurina:2021lwr,Badurina:2022ngn}. The upper left panel shows the potential sensitivity to an ULDM-electron coupling $d_{m_e}$ of a 100 m AI, and the upper right panel shows its potential sensitivity to an ULDM-photon coupling $d_e$. The irreducible background due to Atom Shot Noise (ASN) is shown as a black dashed line, and the solid orange and blue lines in this and other panels indicate the GGN background calculated using the New High-Noise Model (NHNM) and the New Low-Noise Model (NLNM)~\cite{peterson1993observations} that are based on extensive surveys of noise levels at different sites, assuming that the surrounding rock is an isotropic layer of sandstone (molasse). The violet line in this and the two right panels is the level of GGN estimated on the basis of seismic measurements at the top and bottom of the PX46 shaft, again assuming that the surrounding rock is sandstone (molasse). These GGN curves are not substantially altered if the rock surrounding PX46 is assumed to be glacial till (moraine).~\footnote{However, PX46 is known to be surrounded by two strata, and the geological environment is anisotropic, so more complete calculations and seismic measurements will be needed to refine the illustrative GGN curves in Fig.~\ref{fig:AI100Sensitivities}.}

\section{Overview of AICE and of the required infrastructure}

{\bf Laser Laboratory}: The laser laboratory located in the surface building SX4 will house the specialised instruments needed for laser cooling and manipulation of atoms in the detector, e.g., lasers, optics, control electronics, computers, and coil drivers. It should have a floor area of at least 50m$^2$, allowing space for these instruments and for testing of each atom source before installation into the 100m detector.

{\bf Laser Link}: Laser light and all data and electrical signals will be delivered from the laser laboratory to the detector through optical fibres and cables. The route for the optical fibres should be kept fairly short ($< 50$m), and isolated from acoustically noisy equipment, to limit optical absorption loss, phase noise and stimulated Brillouin scattering.

{\bf Interferometry region}: The atom interferometry region consists of an ultra-high-vacuum tube with internal length $\sim 100$m and clear diameter of at least 150mm. The vacuum tube is wrapped in field coils and surrounded by multi-layer magnetic shields, and has up to ten connection nodes to which side-arms will be attached. The specifications for services are relatively light because most of the interferometry region is passive~\cite{Arduini:2851946}, but some services are required, e.g., for environmental monitoring as listed in Table~\ref{tab:tech-reqs}.

{{\bf Support structure}: The vacuum tube will be surrounded by a support structure attached to the wall of PX46, as illustrated in the right panel of Fig.~\ref{fig:PX46}. A detailed design of the corresponding structure for the AION-10 experiment was presented in its TDR~\cite{Bongs:2025rqe}, together with a stability analysis. Also, the MAGIS collaboration has designed a support structure for the planned 100-m AI experiment at Fermilab~\cite{MAGIS-100:2021etm}. The design of the support structure for the proposed 100-m AI in PX46 will build upon these studies and draw upon experience with construction of the 17m AI tower in Hannover~\cite{schlippert2020matter}.}

{\bf Atom sources}: The atom sources are attached to the interferometry region as ``side-arms”. In these side-arms, a hot atomic beam is slowed, captured, and laser-cooled, before being transported horizontally into the interferometry region. Each side-arm consists of an ultra-high-vacuum chamber surrounded by field coils, optics, cameras, and local lasers (e.g., 461nm laser diodes). These elements, combined with their support structure and enclosure, occupy a volume of approximately $1 \times 1 \times 2$m$^3$ with a mass of order 200 kg. In addition, each side-arm requires local control electronics for sequence coordination, data acquisition, lasers, and active optics. The service specifications for each side-arm are listed in Table~\ref{tab:tech-reqs}.
The initial configuration may have only two side-arms, but it is anticipated that ultimately 5 to 10 side-arms will be required. The cold-atom technology will require daily intervention such as optical alignment in order to maintain reliable operation. To support these interventions, access to the laser laboratory and side-arms is required for at least 12 hours per day. The access method must be practical for daily use (especially during initial commissioning), and must be safe from radiation, oxygen deficiency and fire hazards. 

{\bf Mirror platform}: The interferometry beam is sent downwards from the top of the interferometry region, and must be retro-reflected by a mirror at the bottom of the tube in order to allow both upward and downward photon momentum kicks to be imparted to the atoms. The mirror must be in-vacuum, and installed on a piezo-tunable platform to allow for dynamic compensation of the rotation of the earth. For the purposes of the CFS~\cite{Arduini:2851946}, the service specifications for the mirror platform were approximated as being the same as for a side-arm, though in practice the requirements for the mirror platform will be less extensive.

\begin{table}[htbp]
\small
\centering
\caption{\label{tab:tech-reqs} Preliminary technical and infrastructure requirements. It is anticipated that 5 to 10 side-arms will ultimately be required.}
\smallskip
\bgroup
\def\arraystretch{1.25}
 \begin{tabular}{|>{\centering\arraybackslash}m{2.5cm}|>{\centering\arraybackslash}m{4cm}|>{\centering\arraybackslash}m{4cm}|>{\centering\arraybackslash}m{4cm}|}
 \hline
 Requirement & Laser Lab & Interferometry region & Side-arm (per side-arm)\\ \hline
 \hline
  Volume & Floor area $>$ \SI{50}{m^2} & \SI{1}{m^2} cross-sectional area &  1~m $\times$ 1~m $\times$ 2~m \\ \hline
  Mains power & $\sim$ \SI{35}{kW} (three- and single-phase outlets) & $\mathcal{O}(\SI{100}{W})$ diagnostic and monitoring electronics & $\mathcal{O}(\SI{10}{kW})$\\ \hline
  Control cables & Ethernet, fibre, coaxial & Magnetic coils, diagnostic and monitoring electronics & optical fibres, coaxial, high-power steel-clad fibers\\ \hline
  Temperature stability & \SI{22}{\degreeCelsius} w/ $\pm$ \SI{1}{\degreeCelsius} pk-pk & $<\SI{1}{\degreeCelsius\per\hour}$ & Temperature controlled, NEMA rated enclosure, $<\SI{0.5}{\degreeCelsius}$ pk-pk\\ \hline
  Water cooling & \SI{30}{kW} cooling capacity & n/a & \SI{5}{kW} cooling capacity, $<\pm \SI{1}{\degreeCelsius}$ stability\\ \hline
   Laser safety & Engineering (enclosures, interlocks); admin (training); PPE (glasses) & Already safe (enclosed) & Engineering (enclosures); admin (training); PPE (glasses) \\ \hline
  Gases & Helium, compressed air, Argon & n/a & Helium for commissioning\\ \hline
  Cryogenics & n/a & n/a & n/a \\ \hline
  Ventilation & Air-handling unit capable of temp. spec. & Air-flow to maintain temp. spec. & Air-flow to move \SI{5}{kW} of heat\\ \hline
  Access & Year-round ($>$ 12 hrs/day) & Access for maintenance (more access during calibration and commissioning) & Year-round $\sim$ 12 hrs/day (more R\& D for fully autonomous atom sources)\\ \hline
  Smoke detector & Yes & Yes & Yes\\ \hline
  Oxygen depletion monitor & Yes & During maintenance & n/a\\ \hline
  Hoisting \newline equipment & n/a & Modular sections $<$ \SI{907}{kg} & n/a \\
 \hline
 \end{tabular}
 \label{tab:tech-reqs}
\egroup
\end{table}

{\bf Environmental systematics}: These include ambient seismic activity and atmospheric temperature and pressure fluctuations, as well as anthropogenic electromagnetic interference. Density perturbations of the Earth and atmosphere couple directly to the atoms through perturbations of the gravitational field. These effects are known as gravity gradient noise (GGN) and generate an important noise floor for long-baseline terrestrial AI. Measurements at the CERN site~\cite{Arduini:2851946} indicate that the amplitude of the seismic spectral density in the target frequency band is intermediate between the NHNM and NLNM~\cite{peterson1993observations}. We also envisage the possibility of implementing active mitigation of GGN using purpose-built instrumentation for seismic and weather monitoring.

{\bf Ambient magnetic and RF fields}: These place requirements on the necessary levels of magnetic and RF shielding. Most static magnetic fields are not of great concern, as the experiment will use active bias fields and magnetic shielding to control the field uniformity. Time-varying magnetic fields around the laser laboratory and in the shaft around the interferometry region and side-arms should be limited to
$\delta B \le 100$ nT/$\sqrt{\rm Hz}$, though magnetic field fluctuations below $100$ pT/$\sqrt{\rm Hz}$ would
be desirable. These constraints apply within the peak detector sensitivity band 50 mHz to 10 Hz, in particular, though strong field noise peaks at Fourier frequencies outside this band could potentially alias into the detector band. Drift or steps in magnetic field of $\gtrsim 50$ nT are also a potential concern for the side-arms and the detector, due to their effect on the Magneto-Optical Trap (MOT) position and the interferometer transition, but slow changes of background magnetic field of up to several $\mu$T could straightforwardly be mitigated by shielding and active field control if required. Measurements of the electromagnetic field noise in PX46 have found that it is acceptably low during LHC operation as well as when it is not operating. The mean magnetic field level changes during the LHC machine cycle, but the rate of change is slow and can be accommodated by the experiment's magnetic shielding, which should be synchronised with LHC operations.

\begin{figure}[h!]
    \centering
    \includegraphics[width=0.49\textwidth]{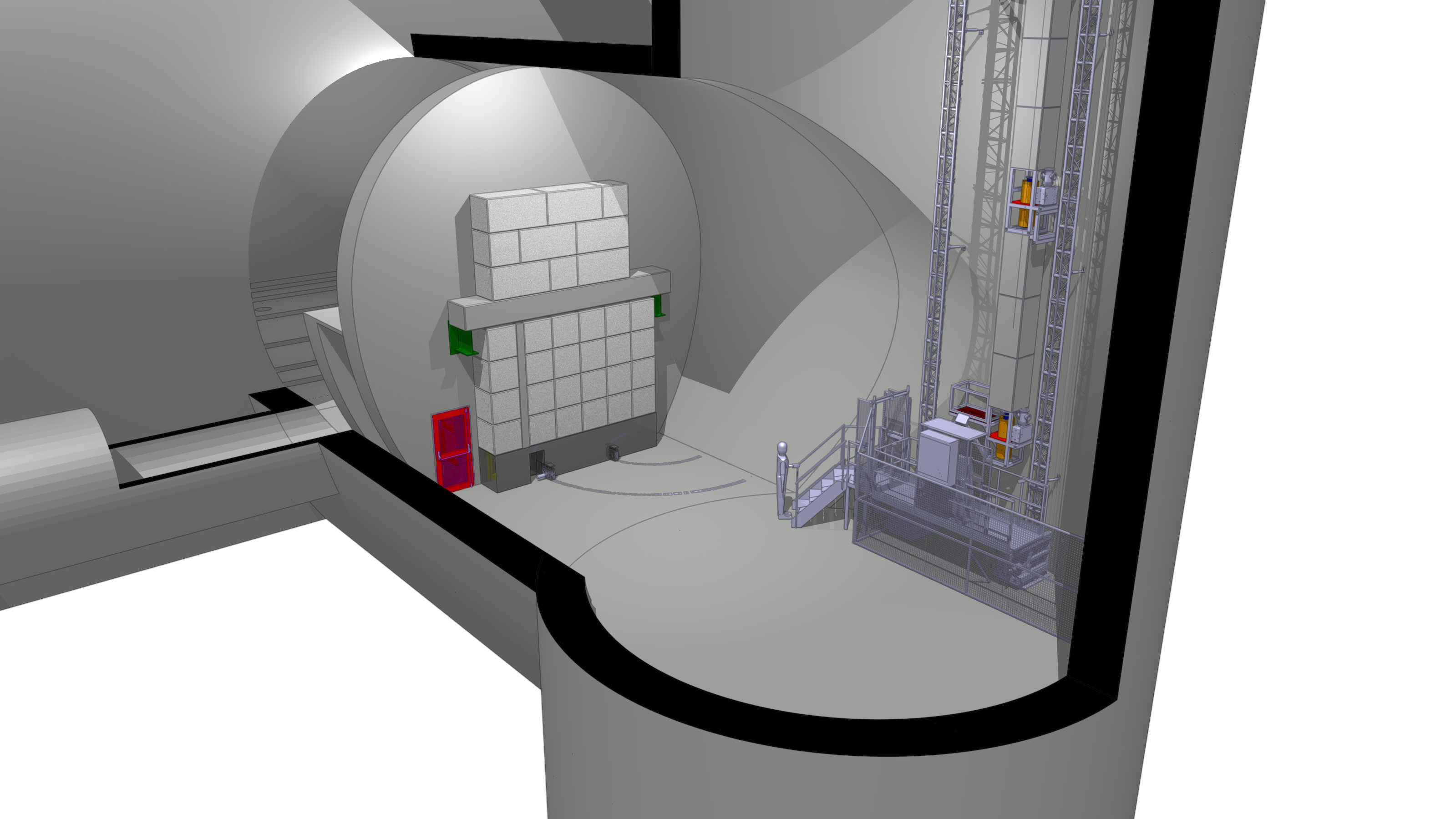}
    \includegraphics[width=0.49\textwidth]{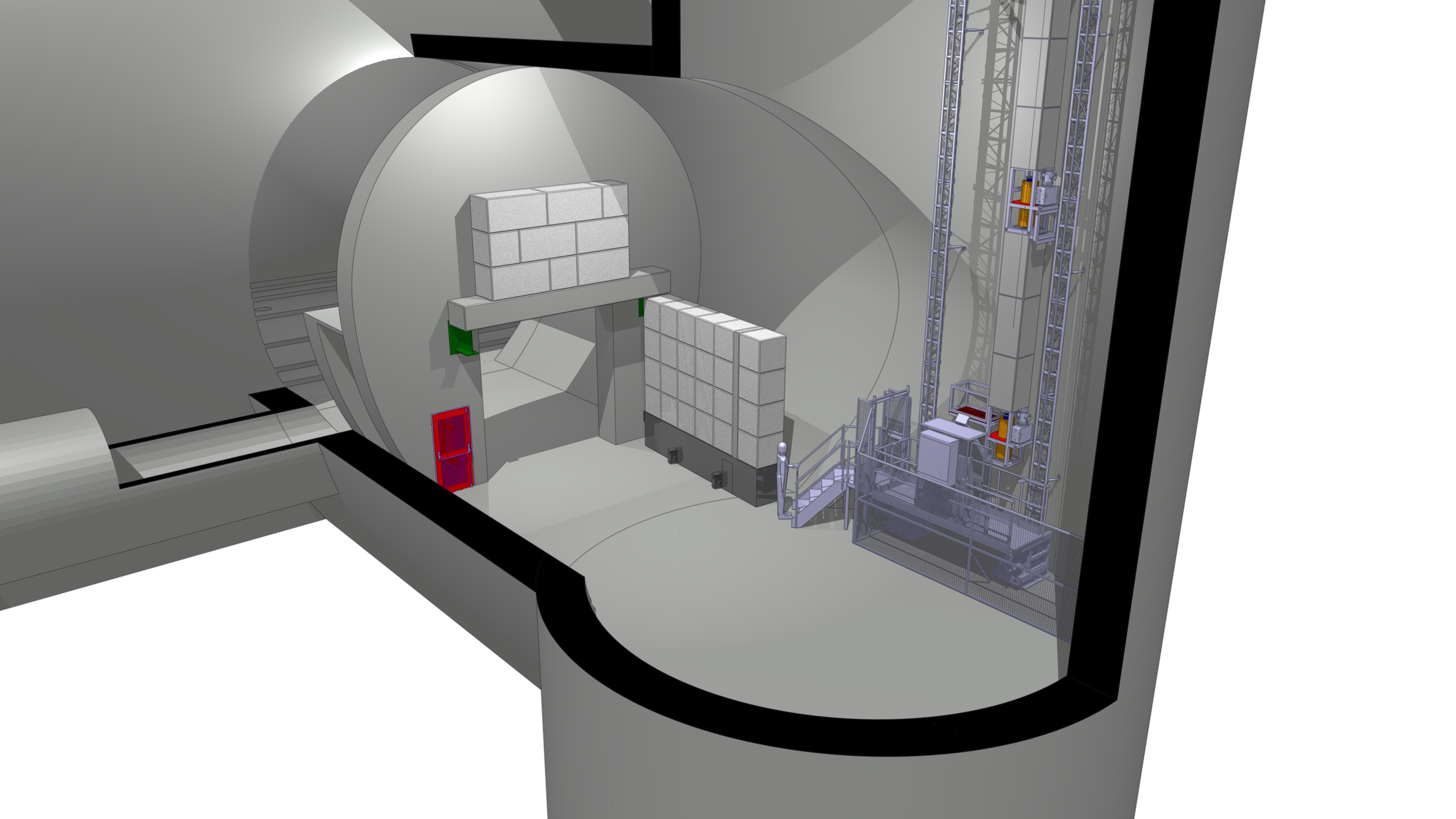}
    \caption{\label{fig:CE:chicane}Shielding wall with the proposed access doors in closed ({\it Left panel}) and open ({\it Right panel}) configuration. The bottom landing of the elevator platform is also visible.}
\end{figure}

{\bf Radiation protection}: Several Radiation Protection (RP) aspects have been taken into account, including shielding requirements, radiation levels during operation (stray or prompt radiation) and technical stops (residual radiation), area classification, radiation monitoring, and more~\cite{Forkel-Wirth:2002}. Estimating the stray radiation field has been crucial for drafting the project proposal, identifying possible showstoppers and in general verifying that the proposal is consistent with CERN RP rules. PX46 is considered a low-occupancy (i.e., $< 20$\% working time) supervised radiation area~\cite{EDMS-810149:2007} accessible only when the LHC is not in operation.
The preliminary RP study conducted for an AI experiment is summarized in~\cite{Maietta:2020}. It was found that, in the absence of additional shielding, a recommended maximum depth of 80m from the surface should be specified and a maximum depth of 90m depth should not be exceeded.
A detailed study has been made of a shielding wall in TX46 (see Fig.~\ref{fig:CE:chicane}), which would allow the useful depth of PX46 to be extended down to the floor. In order to provide a safe evacuation path through the shielding wall, it is proposed to install two new access doors in the shielding wall in TX46: an end-of-zone door and a second ventilation door with an additional 0.8m thick chicane wall. Simulation shows that the proposed solution would allow the possibility of accessing the full 143m depth of the PX46 shaft during LHC operation, which will remain a  supervised radiation area while guaranteeing operators' safety in in the case of an accidental beam loss event.

{\bf Fire and helium release}: These safety risks have been evaluated and mitigation measures have been devised. From a general point of view, in the event of fire (smoke) or helium hazard, escape of the operators of the AI must be possible in the shortest time and safest way possible. This led to the choice of installing a single elevator platform that can be used for normal access and for escape in all cases of emergencies, and which complies with all relevant requirements, regulations and rules~\cite{Arduini:2851946}. Specific auxiliary means of access for the fire brigade rescue teams will also be implemented, in order to allow the occupants to be reached and evacuated if necessary.
Two major scenarios of uncontrolled helium release giving rise to a safety hazard are possible at LHC point 4: either a release from the superconducting RF cryomodules located in the RUX45 section of the tunnel or a release from the main LHC superconducting dipole strings in the arcs 34 or 45. These scenarios have been considered in light of experience with the helium release from the RF cryomodules in August 2022, when measured oxygen levels in the TU46 ventilation gallery remained within acceptable limits, and the helium release from the LHC dipole string in September 2008, following which confinement doors in the LHC tunnel have been installed or upgraded in order to route safely the helium flow. The AI experiment would itself be further separated by the planned radiation shielding wall in the TX46 gallery. Moreover, it is anticipated that the AI operators would carry self-rescue masks while accessing the experiment, helping to minimize the risk.

\begin{figure}[h!]
	\centering
    \includegraphics[width=0.45\textwidth]{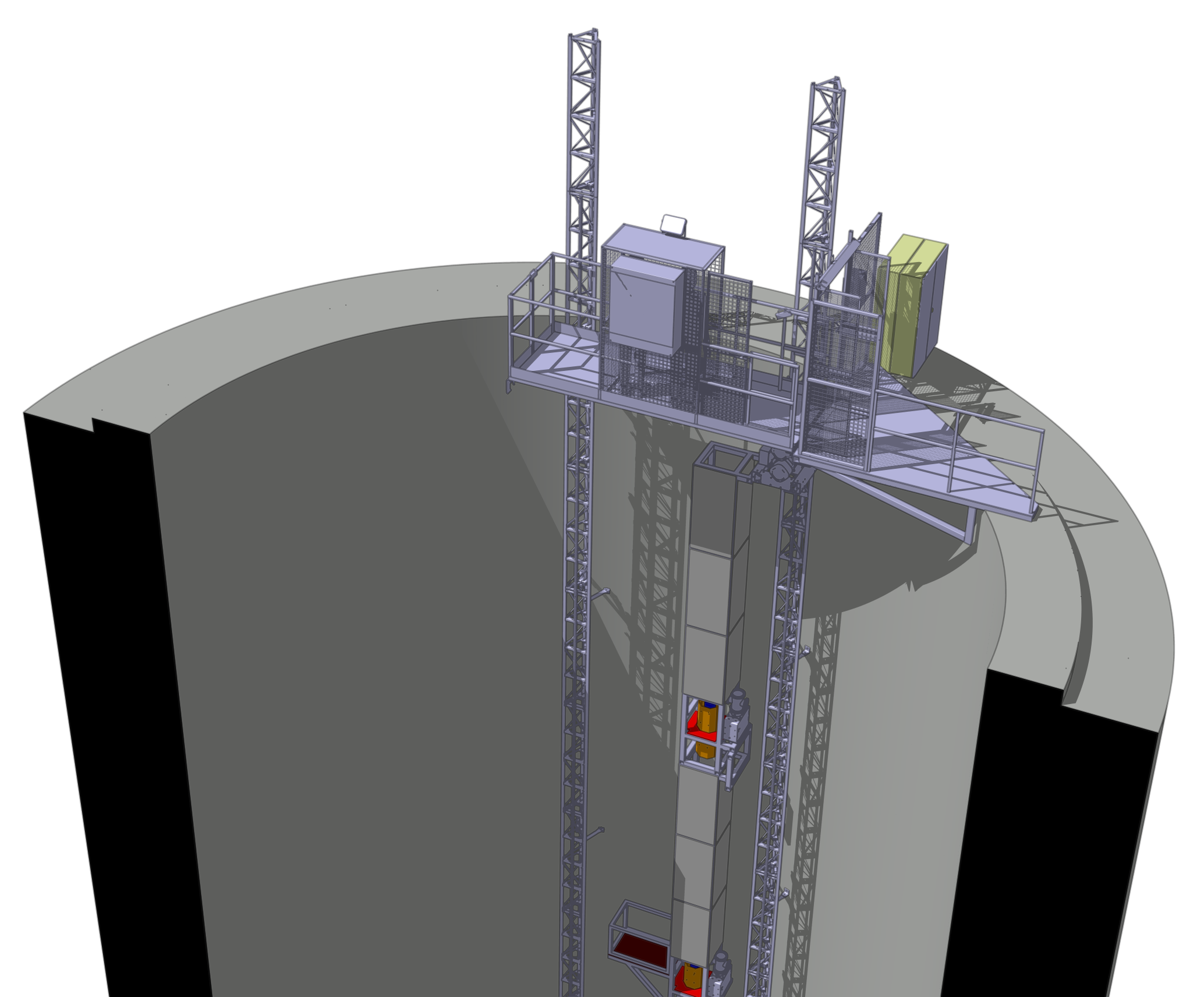}
    \includegraphics[width=0.45\textwidth]{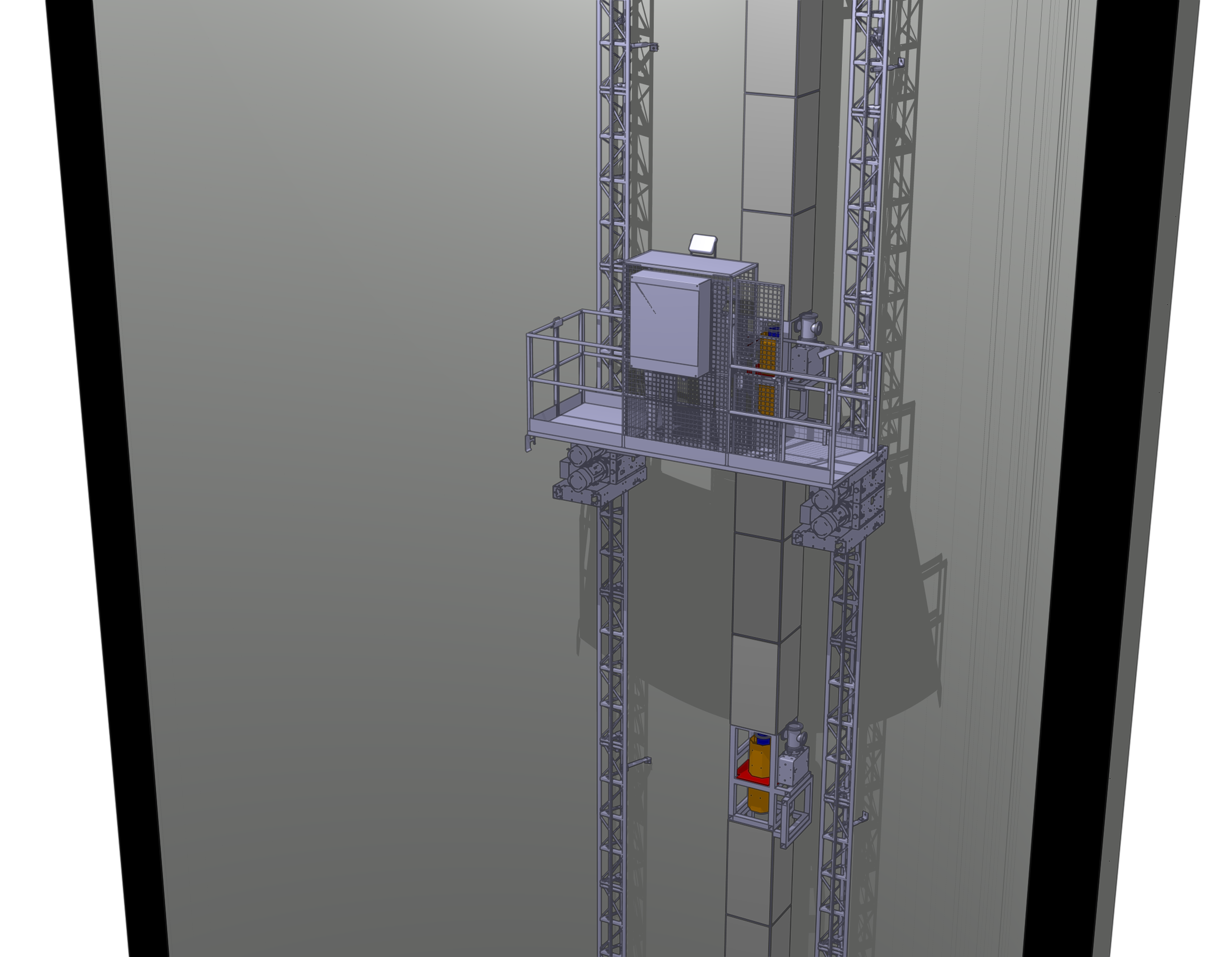}
	\caption{\label{fig:Platform1} 3D view {\it (Left panel)} of the top landing, and {\it (Right panel)} of the dedicated elevator platform in PX46 during operation~\cite{XLIndustries:2023}.}
\end{figure}

{\bf Elevator platform}: If a fire is detected in PX46 or one of the neighboring areas (TX46, UX45), personnel must evacuate using the elevator platform that is used to access the AI and reach the top of the shaft or, should evacuation to the surface not be possible, reaching the bottom of the shaft. The proposed platform (see Fig.~\ref{fig:Platform1}) is electrically powered and has backup batteries. If these batteries fail, the platform has been designed to be able to perform a controlled mechanically braked descent to the bottom of the shaft in less than 2 minutes. AI operators should then evacuate through the access door at the bottom of PX46 using one of the existing LHC escape routes~\cite{LHC-0000006238:2013}. The 2-minute evacuation time as been assessed in~\cite{Arduini:2851946} and does not introduce a significant risk for the AI personnel.

{\bf Surface access}: Currently the top of the shaft is closed with a steel cover that can be opened in the middle with a hoist to allow transport into and out of the shaft. This will require some modification of the design to allow the installation of the experiment with the proposed elevator platform and regular access from the top. A new enclosed access area will be created for this purpose with an airlock  controlled by a badge reader connected to the LHC Access Control System.

{\bf Access control system}: Modifications to the access control system consist of two distinct parts: access control to the shaft PX46, which will normally be classified as a supervised radiation area, and access control to the LHC machine interlocked via the LASS through the door located in the shielding wall. These have been studied in detail, and the proposed solutions have been costed.

{\bf Heating, ventilation and air conditioning (HVAC)}: The access doors illustrated in Fig.~\ref{fig:CE:chicane} have a meshed insert. This stems from the requirement of having some airflow at the bottom of PX46 when the shielding wall is constructed in order to avoid stagnation and accumulation of CO2 creating a potential hazard for the AI personnel. Meshed doors comply with this requirement, while not adding any extra impact in case of fire or of an helium release accident. The airlock at the top allows maintaining a slight overpressure inside the PX46 shaft guaranteeing normal LHC ventilation from the equipment located in the SX46 building. 
No further activities will be required in the HVAC domain during the realization of the preparation of PX46 to host an AI experiment.  

\section{Infrastructure Cost Drivers and Schedule Constraints}

When assessing the costs associated with preparing PX46 for the installation of a 100m AI experiment, we  distinguish between the generic costs associated with such an experiment - such as the establishment of a laser laboratory, the support framework for the vacuum tube and electricity distribution - from the specific infrastructure costs associated with installing such an experiment at CERN. The latter include the measures necessary for protection against radiation and fire hazards, such as a shielding wall and a suitable elevator platform, and to control access to the AI experimental area.

As discussed above, a key requirement is to provide radiation protection for people working on the experiment, even in the event of a catastrophic LHC beam loss close to the RF system at the base of the PX46 shaft. This requires the installation of a shielding wall that is configured to permit the transportation of LHC machine components while offering robust radiation protection. As described above, it is proposed to locate the shielding wall in the TX46 access gallery.

Another key requirement for safe operation at CERN is provision for evacuation from PX46 within $\sim 2$ minutes in the event of a fire. This time restriction is incompatible with evacuation via stairs or a conventional (relatively slow-moving) lift. Consulting engineers have proposed a suitable elevator platform.

Access to PX46, like other CERN areas and facilities, will be subject to restrictions enforced by the LHC Access and Safety Control systems and monitored by adequate safety systems (smoke detection, ODH alarms, etc.).

\begin{table}[htbp]
\centering
\caption{Class 4 budget estimates for civil engineering, access and alarm system modifications, lifting platform and other heavy handling and transport equipment to be installed during LS3.}
\bigskip
\begin{tabular}{|l|c|}
\hline
{\bf Description}                                    & {\bf Cost (CHF)} \\
\hline
Civil engineering design services                    & 130'000 \\
Civil engineering works (including concrete blocks and door rails)  & 260'000 \\
LHC Access Safety System (LASS)                      & 65'000 \\
Fire detection, alarms, emergency communications     & 30'000  \\
RP monitor                                           & 20'000  \\
Lifting platform                                     & 400'000  \\
Movable shielding door (structure)                   & 100'000  \\
Modifications of shaft top plug (including ventilation room)                   & 75'000 \\
Additional hoist in SX4                              & 130'000 \\
\hline
{\bf Grand total}                                    & {\bf 1'210'000} \\
\hline
\end{tabular}
\label{tab:total_costs}
\end{table}

The main cost items for the infrastructure required for the preparation of PX46 for the installation of an AI during LHC Run~4 are listed in Table~\ref{tab:total_costs}. These include civil engineering, access and safety systems, the elevator platform, a movable shielding door, modifications of the shaft top plug and an additional hoist in SX4. It is anticipated that the laser laboratory (with the associated access interlock system) and the AI with its ancillaries will be entirely provided by the experimental collaboration. This is significantly smaller than the cost of the experiment itself which is estimated to be in the range 30 to 50 MCHF.

\begin{figure}
    \centering
    \includegraphics[width=1.0\linewidth]{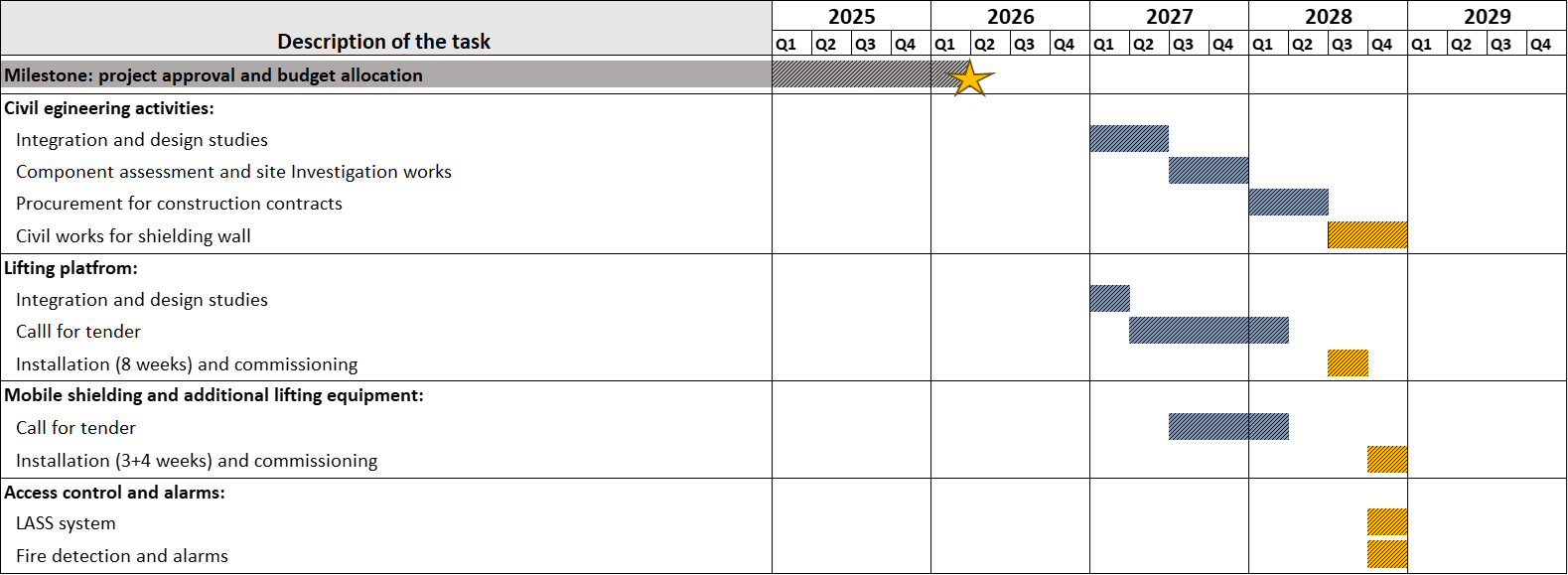}
    \caption{Gantt chart outlining the timeline of the major work items described in this report. Preparatory activities are indicated in gray, while field work is indicated in orange.}
    \label{fig:Gantt}
\end{figure}

A preliminary schedule for the realization of all the preparatory work described in this report has been developed~\cite{Arduini:2025jhe}, and is illustrated in Gantt form in Fig.~\ref{fig:Gantt}.
The schedule is subject to a few constraints. Obviously, timely approval of the project and its budget is necessary. Civil engineering works must be completed before installation of the movable shielding door and of the personnel access doors, and before instrumenting them and interfacing with the LASS. The tendering phase of the platform, of the mobile shielding and other lifting equipment has a rather long lead time. Moreover, the relevant technical services have specific internal time frames for the installation work, preferentially in the second half of 2028. Finally, it is not feasible to install fire detection in the shaft before the lifting platform is in place. These constraints are mostly logistical and do not actually introduce any significant delays on a technically limited schedule. 



\section{Conclusions}

A 100m AI experiment would have unique capabilities to search for ultralight dark matter, and would open the way to searches for gravitational waves and other interesting phenomena in fundamental physics. Feasibility and implementation studies have not identified any showstoppers for the installation and operation of such an experiment in the PX46 shaft at CERN. It would complement the existing CERN programme in a novel way and providing a unique addition to the Physics Beyond Colliders portfolio that aligns very well with CERN’s Quantum Technology Initiative.

The feasibility~\cite{Arduini:2851946} and implementation~\cite{Arduini:2025jhe} studies have shown that AICE, its infrastructure and environmental constraints could be accommodated at CERN without interfering with the LHC scientific programme. Exploratory measurements indicate that the experimental requirements for the spectra of vibrations, seismic and electromagnetic noise could all be met in PX46. Health and safety issues specific to CERN such as radiation protection, fire safety and helium release incidents have all been considered and suitable technical solutions found. Essential aspects of the proposed new infrastructure were investigated, including civil engineering, experimental access and the emergency evacuation of personnel, access control, ventilation, cooling and electricity supply.

The extra preparatory costs associated with installing a 100m atom interferometer at CERN rather than elsewhere (see Table~\ref{tab:total_costs}) would be small compared with the overall cost of such an experiment.

We anticipate that, if the CERN management encourages this LoI, a significant fraction of Terrestrial Very Long Baseline Atom Interferometer (TVLBAI) Proto-Collaboration~\cite{abend_terrestrial_2024,abdalla_terrestrial_2025,TVLBAIESPP} would wish to participate in AICE (see Appendix A).

\bibliographystyle{JHEP}
\bibliography{CERNAI}

\providecommand{\href}[2]{#2}\begingroup\raggedright\begin{thebibliography}{10}

\bibitem{Arduini:2851946}
G.~Arduini et~al., \emph{{A Long-Baseline Atom Interferometer at CERN:
  Conceptual Feasibility Study}},  Tech. Rep.
  \href{https://cds.cern.ch/record/2851946}{CERN-PBC-REPORT-2023-002,
  CERN-TH-2023-051}, CERN, Geneva (2023).

\bibitem{Arduini:2025jhe}
G.~Arduini et~al., \emph{{A Long-Baseline Atom Interferometer at CERN LHC Point
  4: Implementation Study}}, {\emph{CERN-PBC Report-2025-004} (2025) }
  [\href{https://arxiv.org/abs/2508.09694}{{\ttfamily 2508.09694}}].

\bibitem{abend_terrestrial_2024}
S.~Abend et~al., \emph{Terrestrial very-long-baseline atom interferometry:
  {Workshop} summary}, \href{https://doi.org/10.1116/5.0185291}{\emph{AVS
  Quantum Science} {\bfseries 6} (2024) 024701}.

\bibitem{abdalla_terrestrial_2025}
A.~Abdalla et~al., \emph{Terrestrial {Very}-{Long}-{Baseline} {Atom}
  {Interferometry}: summary of the second workshop},
  \href{https://doi.org/10.1140/epjqt/s40507-025-00344-3}{\emph{EPJ Quantum
  Technology} {\bfseries 12} (2025) 42}.

\bibitem{TVLBAI3}
\emph{3rd {Terrestrial} {Very}-{Long}-{Baseline} {Atom} {Interferometry}
  {Workshop}},  {Leibniz} {University} - {Hannover},
  \href{https://indico.cern.ch/event/1522217/}{https://indico.cern.ch/event/1522217/}.

\bibitem{TVLBAIESPP}
{\scshape TVLBAI Proto} collaboration, \emph{{Long-Baseline Atom Interferometry
  (input to ESPP 2026 update)}},
  \href{https://arxiv.org/abs/2503.21366}{{\ttfamily 2503.21366}}.

\bibitem{Badurina:2019hst}
L.~Badurina et~al., \emph{{AION: An Atom Interferometer Observatory and
  Network}}, \href{https://doi.org/10.1088/1475-7516/2020/05/011}{\emph{JCAP}
  {\bfseries 05} (2020) 011}
  [\href{https://arxiv.org/abs/1911.11755}{{\ttfamily 1911.11755}}].

\bibitem{schlippert2020matter}
D.~Schlippert et~al., \emph{Matter-wave interferometry for inertial sensing and
  tests of fundamental physics},  in \emph{CPT AND LORENTZ SYMMETRY:
  Proceedings of the Eighth Meeting on CPT and Lorentz Symmetry}, pp.~37--40,
  World Scientific, 2020.

\bibitem{Overstreet:2021hea}
C.~Overstreet, P.~Asenbaum, J.~Curti, M.~Kim and M.A.~Kasevich,
  \emph{{Observation of a gravitational Aharonov-Bohm effect}},
  \href{https://doi.org/10.1126/science.abl7152}{\emph{Science} {\bfseries 375}
  (2021) abl7152}.

\bibitem{zhou2011development}
L.~Zhou, Z.~Xiong, W.~Yang, B.~Tang, W.~Peng, K.~Hao et~al., \emph{Development
  of an atom gravimeter and status of the 10-meter atom interferometer for
  precision gravity measurement}, {\emph{General Relativity and Gravitation}
  {\bfseries 43} (2011) 1931}.

\bibitem{Bongs:2025rqe}
K.~Bongs et~al., \emph{{AION-10: Technical Design Report for a 10m Atom
  Interferometer in Oxford}},
  \href{https://arxiv.org/abs/2508.03491}{{\ttfamily 2508.03491}}.

\bibitem{MAGIS-100:2021etm}
{\scshape MAGIS-100} collaboration, \emph{{Matter-wave Atomic Gradiometer
  Interferometric Sensor (MAGIS-100)}},
  \href{https://doi.org/10.1088/2058-9565/abf719}{\emph{Quantum Sci. Technol.}
  {\bfseries 6} (2021) 044003}
  [\href{https://arxiv.org/abs/2104.02835}{{\ttfamily 2104.02835}}].

\bibitem{Canuel:2017rrp}
B.~Canuel et~al., \emph{{Exploring gravity with the MIGA large scale atom
  interferometer}},
  \href{https://doi.org/10.1038/s41598-018-32165-z}{\emph{Sci. Rep.} {\bfseries
  8} (2018) 14064} [\href{https://arxiv.org/abs/1703.02490}{{\ttfamily
  1703.02490}}].

\bibitem{LIGOScientific:2014pky}
{\scshape LIGO Scientific} collaboration, \emph{{Advanced LIGO}},
  \href{https://doi.org/10.1088/0264-9381/32/7/074001}{\emph{Class. Quant.
  Grav.} {\bfseries 32} (2015) 074001}
  [\href{https://arxiv.org/abs/1411.4547}{{\ttfamily 1411.4547}}].

\bibitem{VIRGO:2014yos}
{\scshape VIRGO} collaboration, \emph{{Advanced Virgo: a second-generation
  interferometric gravitational wave detector}},
  \href{https://doi.org/10.1088/0264-9381/32/2/024001}{\emph{Class. Quant.
  Grav.} {\bfseries 32} (2015) 024001}
  [\href{https://arxiv.org/abs/1408.3978}{{\ttfamily 1408.3978}}].

\bibitem{Aso:2013eba}
{\scshape KAGRA} collaboration, \emph{{Interferometer design of the KAGRA
  gravitational wave detector}},
  \href{https://doi.org/10.1103/PhysRevD.88.043007}{\emph{Phys. Rev. D}
  {\bfseries 88} (2013) 043007}
  [\href{https://arxiv.org/abs/1306.6747}{{\ttfamily 1306.6747}}].

\bibitem{Gaignant:2021}
{Ch. Gaignant, S. Weisz}, ``{Doors for LHC operation: Control and release of
  Helium, radiation and radioactive air management}.''
  \href{https://edms.cern.ch/document/1346139/1.4}{LHC-SQ-EY-0001 v.1.4}, 2021.

\bibitem{Hakulinen:2022}
{T. Hakulinen, N. Broca}, ``{Helium venting in P4: ODH levels and alarms logged
  during the event}.'' \href{https://edms.cern.ch/document/2797089/1}{LHC CSAP
  \#{}52 10.11.2022}, 2022.

\bibitem{XLIndustries:2023}
XLIndustries, ``{CERN AION-100 Plateforme de travail}.'' {Confidential}, 2023.

\bibitem{LTS:ACC-PM-MS}
{Jean-Philippe Tock et al.}, \emph{Long-term schedule for the cern accelerator
  complex},  2025.

\bibitem{Bertone:2016nfn}
G.~Bertone and D.~Hooper, \emph{{History of dark matter}},
  \href{https://doi.org/10.1103/RevModPhys.90.045002}{\emph{Rev. Mod. Phys.}
  {\bfseries 90} (2018) 045002}
  [\href{https://arxiv.org/abs/1605.04909}{{\ttfamily 1605.04909}}].

\bibitem{Graham:2015ifn}
P.W.~Graham, D.E.~Kaplan, J.~Mardon, S.~Rajendran and W.A.~Terrano, \emph{{Dark
  Matter Direct Detection with Accelerometers}},
  \href{https://doi.org/10.1103/PhysRevD.93.075029}{\emph{Phys. Rev. D}
  {\bfseries 93} (2016) 075029}
  [\href{https://arxiv.org/abs/1512.06165}{{\ttfamily 1512.06165}}].

\bibitem{LISA:2017pwj}
{\scshape LISA} collaboration, \emph{{Laser Interferometer Space Antenna}},
  \href{https://arxiv.org/abs/arXiv:1702.00786}{{\ttfamily arXiv:1702.00786}}.

\bibitem{EventHorizonTelescope:2019dse}
{\scshape Event Horizon Telescope} collaboration, \emph{{First M87 Event
  Horizon Telescope Results. I. The Shadow of the Supermassive Black Hole}},
  \href{https://doi.org/10.3847/2041-8213/ab0ec7}{\emph{Astrophys. J. Lett.}
  {\bfseries 875} (2019) L1}
  [\href{https://arxiv.org/abs/1906.11238}{{\ttfamily 1906.11238}}].

\bibitem{EventHorizonTelescope:2022wkp}
{\scshape Event Horizon Telescope} collaboration, \emph{{First Sagittarius A*
  Event Horizon Telescope Results. I. The Shadow of the Supermassive Black Hole
  in the Center of the Milky Way}},
  \href{https://doi.org/10.3847/2041-8213/ac6674}{\emph{Astrophys. J. Lett.}
  {\bfseries 930} (2022) L12}.

\bibitem{AION:2025igp}
{\scshape AION} collaboration, \emph{{A Prototype Atom Interferometer to Detect
  Dark Matter and Gravitational Waves}},
  \href{https://arxiv.org/abs/2504.09158}{{\ttfamily 2504.09158}}.

\bibitem{chaibi2016low}
W.~Chaibi, R.~Geiger, B.~Canuel, A.~Bertoldi, A.~Landragin and P.~Bouyer,
  \emph{Low frequency gravitational wave detection with ground-based atom
  interferometer arrays}, {\emph{Physical Review D} {\bfseries 93} (2016)
  021101}.

\bibitem{Badurina:2022ngn}
L.~Badurina, V.~Gibson, C.~McCabe and J.~Mitchell, \emph{{Ultralight dark
  matter searches at the sub-Hz frontier with atom multigradiometry}},
  \href{https://doi.org/10.1103/PhysRevD.107.055002}{\emph{Phys. Rev. D}
  {\bfseries 107} (2023) 055002}
  [\href{https://arxiv.org/abs/2211.01854}{{\ttfamily 2211.01854}}].

\bibitem{Arvanitaki:2016fyj}
A.~Arvanitaki, P.W.~Graham, J.M.~Hogan, S.~Rajendran and K.~Van~Tilburg,
  \emph{{Search for light scalar dark matter with atomic gravitational wave
  detectors}}, \href{https://doi.org/10.1103/PhysRevD.97.075020}{\emph{Phys.
  Rev. D} {\bfseries 97} (2018) 075020}
  [\href{https://arxiv.org/abs/1606.04541}{{\ttfamily 1606.04541}}].

\bibitem{Badurina:2021lwr}
L.~Badurina, D.~Blas and C.~McCabe, \emph{{Refined ultralight scalar dark
  matter searches with compact atom gradiometers}},
  \href{https://doi.org/10.1103/PhysRevD.105.023006}{\emph{Phys. Rev. D}
  {\bfseries 105} (2022) 023006}
  [\href{https://arxiv.org/abs/2109.10965}{{\ttfamily 2109.10965}}].

\bibitem{peterson1993observations}
J.~Peterson et~al., \emph{Observations and modeling of seismic background
  noise}, vol.~93, US Geological Survey Reston, VA, USA (1993).

\bibitem{Forkel-Wirth:2002}
D.~Forkel-Wirth et~al., \emph{{Radiation protection at CERN}},
  \href{https://arxiv.org/abs/1303.6519}{{\ttfamily 1303.6519}}.

\bibitem{EDMS-810149:2007}
{D. Forkel-Wirth}, ``{Zonage radiologique au CERN}.''
  \href{https://edms.cern.ch/document/810149/1}{EDMS 810149}, 2007.

\bibitem{Maietta:2020}
{A. Infantino and M. Maietta}, ``{Preliminary RP evaluation of the AION
  experiment in LHC Point 4}.''
  \href{https://edms.cern.ch/document/2333747/1.1}{EDMS 2333747}, 2020.

\bibitem{LHC-0000006238:2013}
{CERN~}, ``{II.2.3 - Fire safety}.''
  \href{https://edms.cern.ch/project/LHC-0000006238}{EDMS Node LHC-0000006238},
  2013.

\bibitem{TVLBAIMOU}
``Memorandum of {U}nderstanding for the {T}errestrial {V}ery {L}ong {B}aseline
  {A}tom {I}nterferometer {S}tudy.'' \url{https://indi.to/fshnp}.

\bibitem{TVLBAISIG}
``{S}ignatories of the {M}emorandum of {U}nderstanding for the {T}errestrial
  {V}ery {L}ong {B}aseline {A}tom {I}nterferometer {S}tudy.''
  \url{https://indi.to/ZWn4h}.

\end{thebibliography}\endgroup

\newpage
\appendix

\section{TVLBAI Support and Endorsement}

The 100m atom interferometer experiment outlined in this Letter of Intent is an integral part of the Terrestrial Very Long Baseline Atom Interferometer (TVLBAI) Proto-Collaboration activities and enjoys the full support of the TVLBAI community. The TVLBAI Proto-Collaboration was established through a Memorandum of Understanding that has been signed by 55 institutions in 20 countries with 3 additional observer institutions~\cite{TVLBAIMOU,TVLBAISIG}, representing a coordinated international effort to develop atom interferometers beyond the 10m scale. The scientific objectives and coordinated approach of the TVLBAI community have been developed through three major international workshops, each with over 200 participants from the particle physics, atomic physics, astrophysics and cosmology communities~\cite{abend_terrestrial_2024,abdalla_terrestrial_2025,TVLBAI3}, and the community has provided input to the 2026 update of the European Strategy for Particle Physics~\cite{TVLBAIESPP}.

The proposed experiment at CERN has been identified by the TVLBAI Study Group as a priority demonstrator within their comprehensive roadmap for long-baseline atom interferometry. The TVLBAI activity fully endorses this initiative as a cornerstone activity that directly advances the TVLBAI scientific and technical objectives. As outlined in the TVLBAI roadmap, this experiment represents a key component of the envisioned global network of detectors with baselines $\mathcal{O}(100)$m that will provide unique sensitivity to bosonic ultra-light dark matter couplings and enable first explorations of gravitational wave signals with frequencies $\mathcal{O}(1)$Hz.

As a TVLBAI-supported activity, this experiment benefits from:
\begin{itemize}
\item Coordinated international expertise from the 55 participating TVLBAI institutions across 20 countries;
\item Shared technological development across the global TVLBAI network, including exchange of different approaches using various atomic species (rubidium, strontium, ytterbium) and geometries;
\item Integrated scientific planning as part of the broader TVLBAI demonstrator programme leading towards km-scale detectors in the mid-2030s;
\item Collaborative resource mobilization leveraging the established international framework for information exchange and coordination;
\item Access to the TVLBAI international topical working groups and regular coordination meetings.
\end{itemize}

The TVLBAI Proto-Collaboration recognizes CERN's PX46 site as uniquely promising for demonstrating the feasibility and scientific potential of $\mathcal{O}(100)$m scale atom interferometry. This experiment will serve as a flagship TVLBAI demonstrator, providing crucial validation of technologies and methodologies that will inform the development of the broader TVLBAI network, including future km-scale detectors.

We foresee that a significant fraction of the TVLBAI Proto-Collaboration will actively contribute to and participate in this experiment, bringing together the collective expertise and resources of this established international community to ensure its success. The experiment aligns perfectly with the TVLBAI objectives to reinforce collaboration between researchers, pool resources and expertise, and advance the frontiers of atom interferometry to enable new possibilities in fundamental physics research.

\section{Definitions of acronyms}
\label{sec:Acronyms}

\textbf{}
\setlength{\parindent}{0pt}

\noindent
\textbf{AI}: Atom Interferometer/Interferometry

\textbf{AICE}: Atom Interferometry CERN Experiment

\textbf{AION}: Atom Interferometer Observatory and Network

\textbf{ASN}: Atom Shot Noise

\textbf{AURIGA}: Antenna Ultracriogenica Risonante per l'Indagine Gravitazionale Astronomica~\footnote{Ultracryogenic Resonant Bar Gravitational Wave Detector}

\textbf{BH}: Black Hole

\textbf{BSM}: Beyond the Standard Model

\textbf{CV}: Cooling and Ventilation

\textbf{DM}: Dark Matter

\textbf{EM}: ElectroMagnetic

\textbf{EMC}: Electromagnetic Compatibility

\textbf{GGN}: Gravity Gradient Noise

\textbf{GW}: Gravitational Wave

\textbf{HL-LHC}: High-Luminosity LHC

\textbf{HVAC}; Heating, Ventilation and Air Conditioning

\textbf{IMBH}: Intermediate-Mass Black Holes

\textbf{KAGRA}: KAmioka GRAvitational wave detector

\textbf{LACS}: LHC Access Control System

\textbf{LASS}: LHC Access Safety System

\textbf{LHC}: Large Hadron Collider

\textbf{LIGO}: Laser Interferometer Gravitational Observatory experiment

\textbf{LISA}: Laser Interferometer Space Antenna

\textbf{LMT}: Large Momentum Transfer

\textbf{LS}: Long Shutdown

\textbf{LSBB}: Laboratoire Souterrain {\` a} Bas Bruit~\footnote{Low-Noise Underground Laboratory}

\textbf{LVK}: LIGO, Virgo and KAGRA

\textbf{MAD}: Material Access Device

\textbf{MAGIS}: Matter-wave Atomic Gradiometer Interferometric Sensor experiment

\textbf{MCI}: Maximum Credible Incident

\textbf{MICROSCOPE}: Micro-Satellite à traînée Compensée pour l'Observation du Principe d'Equivalence~\footnote{Micro-Satellite with Compensated Drag for Observing the Principle of Equivalence}

\textbf{MIGA}: Matter wave-laser based Interferometer Gravitation Antenna

\textbf{NHNM}: New High-Noise Model

\textbf{NLNM}: New Low-Noise Model

\textbf{ODH}: Oxygen Deficiency Hazard

\textbf{PAD}: Personal Access Device

\textbf{PBC}: Physics Beyond Colliders

\textbf{PM}: Puit Materiel~\footnote{Access shaft with stairs and lift used for the transfer of equipment}

\textbf{PPE}: Personal Protection Equipment

\textbf{PX}: Puit eXperience~\footnote{Access shaft to experimental cavern for (formerly) LEP or (currently) LHC detectors}

\textbf{PX46}: Access shaft at LHC Point 4

\textbf{RP}: Radiation Protection

\textbf{RF}: Radio Frequency

\textbf{SMBH}: Super Massive Black Holes

\textbf{SM}: Standard Model

\textbf{SUSI}: système de SUrveillance des SItes~\footnote{Site Surveillance system}

\textbf{SU4}: Surface building dedicated to the cooling and ventilation at Point 4

\textbf{SX4}: Surface building on top of the PX46 shaft

\textbf{TETRA}: Terrestrial Trunked Radio, formerly known as Trans-European Trunked Radio

\textbf{TVLBAI}: Terrestrial Very-Long-Baseline Atom Interferometer

\textbf{TX46}: Access gallery at LHC Point 4

\textbf{ULDM}: Ultra-Light Dark Matter

\textbf{UX45}: Experimental cavern at LHC Point 4

\textbf{WIMP}: Weakly Interacting Massive Particle

\textbf{YETS}: Year-End Technical Stop

\textbf{ZAIGA}: Zhaoshan Long-baseline Atom Interferometer Gravitation Antenna
\end{document}